*Preprint of paper accepted at the Information and Software Technology Journal*
https://doi.org/10.1016/j.infsof.2022.107106
*Author version of November 24, 2022*

# Lessons Learned to Improve the UX Practices in Agile Projects Involving Data Science and Process Automation

Bruna Ferreira, Silvio Marques, Marcos Kalinowski, Hélio Lopes, Simone D. J. Barbosa

*Departament of Informatics, Pontifical Catholic University of Rio de Janeiro, Rua Marques de Sao Vicente, 225/410, RDC, Gavea, Rio de Janeiro 22451-900, Rio de Janeiro, Brazil*

---

## Abstract

Context: User-Centered Design (UCD) and Agile methodologies focus on human issues. Nevertheless, agile methodologies focus on contact with contracting customers and generating value for them. Usually, the communication between end users (they use the software and have low decision power) and the agile team is mediated by customers (they have high decision power but do not use the software). However, they do not know the actual problems that end users (may) face in their routine, and they may not be directly affected by software shortcomings. In this context, UX issues are typically identified only after the implementation, during user testing and validation.

Objective: Aiming to improve the understanding and definition of the problem in agile projects, this research investigates the practices and difficulties experienced by agile teams during the development of data science and process automation projects. Also, we analyze the benefits and the teams' perceptions regarding user participation in these projects.

Method: We collected data from four agile teams, in the context of an academia and industry collaboration focusing on delivering data science and process automation solutions. Therefore, we applied a carefully designed questionnaire answered by developers, scrum masters, and UX designers. In total, 18 subjects answered the questionnaire.

Results: From the results, we identify practices used by the teams to de- fine and understand the problem and to represent the solution. The practices most often used are prototypes and meetings with stakeholders. Another practice that helped the team to understand the problem was using Lean Inception (LI) ideation workshops. Also, our results present some specific issues regarding data science projects.

Conclusion: We observed that end-user participation can be critical to understanding and defining the problem. They help to define elements of the domain and barriers in the implementation. We identified a need for approaches that facilitate user-team communication in data science projects to understand the data and its value to the users' routine. We also identified insights about the need for more detailed requirements representations to support the development of data science solutions.

---

## 1. Introduction

User-Centered Design (UCD) and Agile are human-focused methodologies (Brhel et al., 2015). UCD focuses on carrying out activities aimed at developing software that provides the best experience for users (Ogunyemi et al., 2019). Agile, by contrast, focuses on delivering value to stakeholders but does not address user experience (UX) issues (Brhel et al., 2015). In this context, UCD activities have been integrated into Agile Projects to support the user experiences design with the final product (Alyahya and Almughram, 2020). Some benefits of involving end users in software projects are (Bano et al., 2018): improved user satisfaction, better understanding of user requirements, improved communication with users, and increased quality in the final product. According to Ferreira et al. (2021), end users help to identify: opportunities for integration with other software, other actors involved in the process, required information as input for development, and usability problems that would impact their work before deploying the software.

Despite the benefits of involving users and UCD activities in software development, this integration poses several challenges. Zaina et al. (2021) highlighted some of these challenges: the end users do not participate in the early stages to define their goals and needs; the UX issues are not documented, hindering the communication of these requirements; and the team makes decisions based on their experiences, without adopting UX-centric approaches. Bano et al. (2018) cited additional challenges: budget, time and resources limitation; the definition of the "correct" user, who knows the domain; and the lack of motivation of users to participate in UX activities.

Brhel et al. (2015) cited the need for empirical research to investigate the stakeholders' involvement and the artifact-mediated communication in specific contexts from a practical point of view.

This research investigates the practices and difficulties experienced by four agile teams during the development process. Also, we analyze the benefits and the teams' perceptions regarding end-user participation in these projects. The research was conducted in the context of an academia-industry collaboration focusing on delivering data science and process automation solutions. Four research questions guided our analysis:

RQ1. Which practices supported understanding and defining the problem in these types of projects?

RQ2. Which practices supported representing the solutions in these types of projects?

RQ3. Which difficulties were encountered in defining and understanding the problem and in representing solutions in these types of projects?

RQ4. How was the end-user participation in these projects?

In this context, we evaluated three issues regarding UX design: defining the problem space, defining the solution space, and improving user participation. It is noteworthy that the study was performed in a specific context: projects involving data science and automation of routine user processes.

From the results, we identified practices used by the teams to define and understand the problem and represent the solution. Also, our results point to some interesting issues regarding projects involving data analysis and data science activities: domain complexity and uncertainties, evaluation of software value to end users, and communication of requirements to the development team.

This paper is organized as follows: Section 2 briefly introduces User-Centered Design and Agile Development, discussing end-user participation in agile projects. In Section 3, we present the methods used and the research context. Next, in Sections 4 and 5, we present the qualitative and quantitative results, respectively, which are discussed in Section 6. Section 7 presents the validity threats. Finally, in Section 8, we present the conclusions and future work.

## 2. User Centered Design and Agile Development

Agile methods and practices are widely employed in the software industry. The Agile Manifesto defines the following principles: individuals and interactions over processes and tools; working software over comprehensive documentation; customer collaboration over contract negotiation; responding to change over following a plan (Fowler et al., 2001). User-Centered Design (UCD) focuses on the design, development, and evaluation of interactive systems for human use (Ogunyemi et al., 2019). According to ISO 13407 (for Standardization, 2019), UCD consists of four activities (for Standardization, 2019; Almughram and Alyahya, 2017): understanding and specifying the context of use, specifying user and organizational requirements, developing design solutions, and evaluating the design according to the requirements. Adopting user-centered approaches can improve usability, user experience, and support in meeting user needs. Agile focuses on developing useful software with customer value, but it does not focus on User Experience (UX) issues (Brhel et al., 2015). In this context, UCD activities have been integrated into Agile projects to support the development of software with better UX (Alyahya and Almughram, 2020).

Agile projects demand velocity, and user research activities can be complex. In this context, UX issues are not always appropriately performed in agile processes (Schön et al., 2020). Zaina et al. (2021) present some challenges in integrating UX information in agile projects: (i) most of the information integrated into the software considers the end users' *interactions* with the product, but the end users do not participate in the early stages of development to define their *needs and goals*; (ii) the project usually lacks UX documentation – UX issues are handled verbally and not represented in artifacts, making the flow of information more complex; (iii) questions about UX are based mainly on the team's experience and not on UX-specific models, theory, or tools.

Brhel et al. (2015) proposed five principles to integrate Agile development and UCD based on a systematic literature review. The first principle is to separate product discovery and product creation. UCD focuses on exten- sive user requirements analysis and interaction design, and Agile focuses on a detailed planning and project phase. Performing design activities in the early phases is essential to support the designers in creating the solution. However, the authors cite that it is necessary to conduct empirical studies to evaluate the effects of extending the discovery activities. The second prin-ciple concerns iterative and incremental design and development. UCD and Agile promote an incremental and interactive approach by collecting stakeholders' feedback in each iteration. The third principle discusses the parallel interwoven creation tracks. Design and development should be performed in parallel because the design activities in the early phases are not enough to define user requirements and interactions. The fourth principle is about continuous stakeholder involvement. Agile requires continuous involvement with stakeholders, and UCD requires direct end-user participation. Both are human-focused. Finally, the fifth principle discusses Artifact-Mediated Communication. Artifacts document and communicate product and design concepts to all stakeholders. The authors cite the need for empirical research to investigate the stakeholders' involvement and the artifact-mediated communication in specific contexts from a practical point of view.

Incorporating user feedback in agile development processes is not trivial; due to the need for fast deliveries, the team collects feedback from only a few users, and only at the final delivery (Ogunyemi et al., 2019). Also, difficulties in using UX methods lead software engineers to adapt artifacts familiar to them to support UX activities (Zaina et al., 2021). Moreover, the lack of clear procedures makes the team rationalize their experience (Yaman et al., 2020).

### *2.1. End-User Participation in Agile Projects*

The end users' satisfaction with the functionalities and operations of their routine delivered by the software is fundamental for the system's success (Zowghi et al., 2015). In this context, meeting the end users' needs is critical for the development team (Buchan et al., 2017). UCD is a way to improve user participation throughout the development and to support the exploration of the technology use context, to characterize users and processes where the product will be involved (Duque et al., 2019).

Unlike the UCD principles, in Agile, the customer assumes two roles: 1) the person that owns the product and 2) the people who interact and are affected by its use (Law and Lárusdóttir, 2015). However, customers do not experience the problems that occur in the end users' routine. To effectively involve end users in project activities, it is necessary to identify adequate user representatives (Buchan et al., 2017). The term "stakeholders" is generic, and the difference between the customer and the end users is not clear in agile. However, customers are not always end users, and UCD requires direct and unmediated contact with end users.

Participatory Design (PD) (cooperative design) addresses the difference between customers and end users. According to this approach, the users should participate directly in the design activities and decision making (Abelein and Paech, 2015). Kautz (2010) discusses the different roles of customers and end users: the customer has decision power but limited understanding of the users' needs and their routine. Also, the customers may not interact with the software. In contrast, the end users interact with the software and know the routine to integrate the software. Customers order the software and will pay for it, and specify the initial requirements (Kautz, 2011).

Some works discuss the integration of Participatory Design and Agile. Rittenbruch et al. (2002) discuss the integration of Extreme (XP) Programming and PD. They cited some problems in this integration: users are represented by customers, and the intersection between requirements and users' needs is not validated; Field studies are not part of XP processes; Customers communicate with developers focusing mainly on technical issues, the focus on user issues may be missing; Lack of practices to integrate design into the process and offer to customer different solution. They include end users together with customers in the creation and validation of user stories to minimize these problems.

Kensing and Munk-Madsen (1993) discuss that PD fails because users and developers do not understand each other. They propose a model of user-developer communication. In this model, six knowledge areas are es- sential to the design process: relevant structures on users' present work, concrete experiences with users' present work, visions and design propos- als, concrete experience with the new system, an overview of technological options, and concrete experience with technological options. They group the methods of communication in these knowledge areas, to facilitate their selection. Kautz (2011) investigates how users and customers work in practices in agile projects. He discusses that the role of users was represented by customers who were managers and team leaders with operational tasks. Also, he presents a framework that explains PD and how, when, and where in agile development it can be applied in their research: users acted as designers through constant feedback; users participated directly and indirectly in the process, they made comments during presentations, provided their viewpoints and explained their work processes; users had participatory (participating in the development and having some decision power), informative (providing information about needs), and consultive (providing feedback and reviews) functions; customers had an authority about the staff, and the users could make decisions about the project.

Bano and Zowghi (2015) presented a systematic literature review that investigated the relationship between user involvement and system success. They identified the benefits of end-user involvement from different perspectives: psychological, managerial, methodological, cultural, and political. We summarize these benefits in Table 1. Although the benefits are well known, there are some challenges in involving users throughout the software development (Bano et al., 2018): *"the budget and resources, time constraints, users' expertise, insufficient training for users, the lack of motivation for involvement or a negative attitude or behavior toward the new system"*.

Table 1: Benefits of user involvement by Bano and Zowghi (2015)

| Perspective | Benefits |
|---|---|
| Psychological | user satisfaction; user acceptance; users will not resist using a new system in their routine; users will show a positive attitude when using the system; increase perceived relevance to the system and user motivation; a higher degree of trust in the development team; long-term relationships between users, customers, and the development team; |
| Managerial | better communication; developing realistic expectation; management will face less resistance by giving the users a sense of dignity of knowing that they are important to the system; reducing the cost of the system by decreasing the risk of too many changes after implementation; helping in conflict resolution; |
| Methodological | better understanding of user requirements; improved quality of resultant application; improved quality of design decisions; help in overcoming implementation failures |
| Cultural | increased system usage; facilitated knowledge sharing; improved user skills |
| Political | democracy in the workplace |

### *2.2. R&D Lean methodology: Stages in the development process*

The steps performed in these research projects are based on the Lean Research and Development (R&D) methodology (Kalinowski et al., 2020b,a). Some projects evaluated in this research went through all stages, while others performed only some of them. We describe the steps used below. Figure 1 presents the stages of the process.

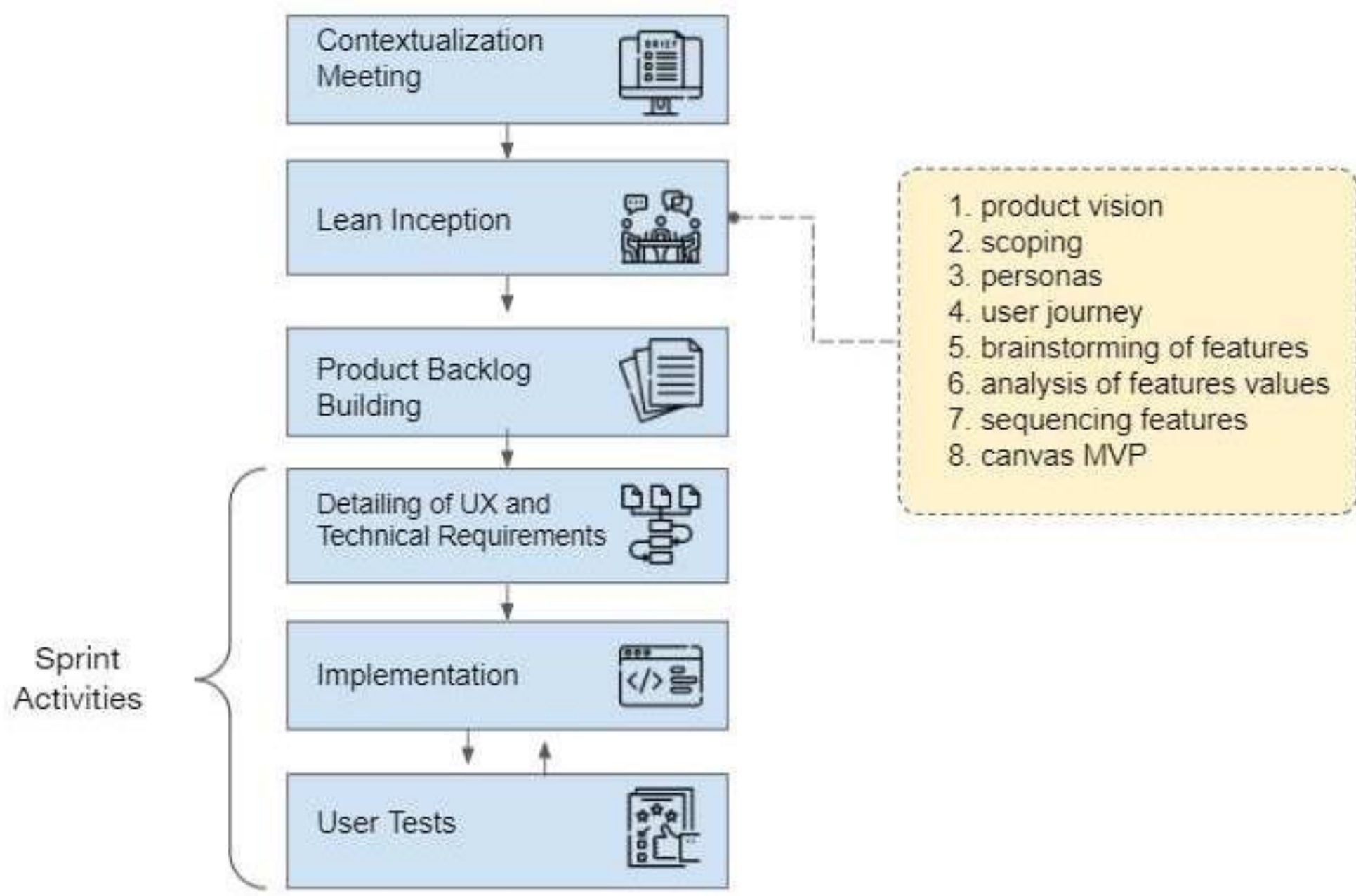

Figure 1: Process followed in the projects and LI activities (Ferreira et al., 2021)

Contextualization: In this stage, customers present to the development team business and technical concepts related to the domain. This stage does not have a specific structure. The stakeholders make presentations using slides. In the end, the team's scrum master makes a brief presentation about Lean Inception (LI), explaining the LI activities as preparation for the LI workshop. Before all the activities of the project, the agile team explains to the customers the important roles involved in the process. In this way, customers are informed about the need to include people that will use the software in their routines –the end users. Also, when the Lean Inception starts, all participants present their roles in an ice breaker session.

Lean Inception: The Lean Inception (LI) steps are based on the guidelines presented in (Caroli, 2018). During the LI, the facilitator and the stakeholders carry out the following activities: description of the product view, delimitation of the product scope (what the product is, what it is not, what it does, and what it does not do), creation of personas to identify the users of the product and their needs, user journey, feature brainstorming, feature value analysis, a sequencer to organize and prioritize product features, and the construction of the Canvas MVP to record the final LI result. The final product in LI is a list of prioritized software features. This stage uses two user-centered methods: personas and user journeys (see templates in Figure 2). In the projects evaluated in this research, the Lean Inception (LI) steps were based on the guidelines presented in (Caroli, 2018) and conducted by a certified facilitator to minimize the bias of compromising results and analyses. The LI was selected together with the customer because it is a workshop that they used in other agile projects that involve innovation. Also, the agile team had previous knowledge about the method application.

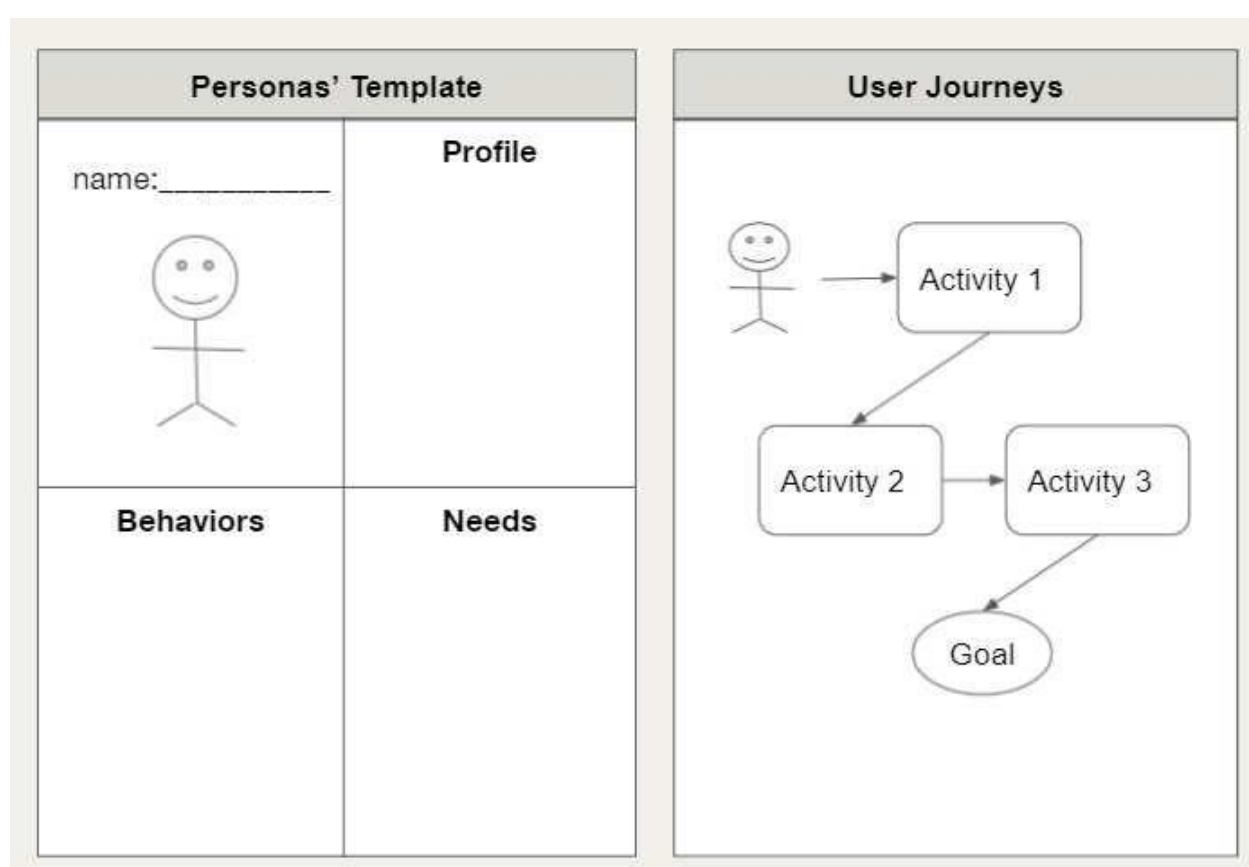

Figure 2: Templates of Personas and User Journeys used in the LI

Product Backlog Building: After the LI, the stakeholders perform a product backlog building (PBB). The procedures of these activities are based on the works of Aguiar and Caroli (2020) and Ferreira et al. (2021). In the first stage of building the backlog, the team and the stakeholders define user needs and problems to be addressed by each feature. Then they list the benefits and expected results and the necessary steps for the development of each feature. The Product Owner guides this activity. Finally, the stakeholders and the agile team analyze the features and create the user stories of the backlog in the format: AS A <user role>, I WANT <functionality> FOR <problem/goal>.

Detailing UX and technical requirements: This step supports the developers and UX/UI designers in understanding the features. All the team, the UX designers, the customers, and the end users participated in this activity. Each project has its characteristics; in this way, the participants of these activities vary. This stage does not have predefined practices and activities. Usually, the facilitator uses the collaborative creation of prototypes to define the solution and its requirements. In each project, the team adopts adequate activities according to their needs and the availability of the participants. In this paper, we explain the practices used by each team in this stage.

Software Implementation and User Tests: In the implementation, the developers coded the software. In some projects, the users participate in this phase to support the team with important information about the process. Finally, in the last stage, the team performs tests with end users and validations with the customers.

## 3. Research Method

We conducted a study to identify practices and difficulties in understanding and defining the problem and representing solutions in projects that aim to support the analysis of data sets in applications involving machine learning, artificial intelligence, and process automation. Also, we investigated the end-user participation in these projects. We used four research questions to guide this study:

RQ1. Which practices supported understanding and defining the problem in these types of projects?

RQ2. Which practices supported representing the solutions in these types of projects?

RQ3. Which difficulties were encountered in defining and understanding the problem and in representing solutions in these types of projects?

RQ4. How was the end-user participation in these projects?

### *3.1. Context Setting and Participants' Profiles*

The evaluated projects are designed and developed in a partnership between academia and industry. The agile teams were organized according to the Lean R&D methodology (Kalinowski et al., 2020b) and were composed of practitioners and some researchers: Scrum masters, product owners, developers, researchers (the academia side) and UX/UI designers (shared across projects), and DevOps Analyst. The academic side was composed of researchers (PhD and Master's students and professors), who conducted the research and supported the agile team in innovation activities. The industry side was from a large publicly held company and was composed of customers with manager profiles, end users, and the technical team responsible for maintaining the software. All the teams worked on an agile co-creation process.

Eighteen members of the agile team participated in this study: two scrum masters, three UX designers/researchers (UX/R), and thirteen de- velopers. These subjects worked in four groups of R&D projects: Digital Twin (DT) and Optimization, Artificial Intelligence Project (AI), Logistic Projects (LOG), and Process Mining Projects (PM). The three UX design- ers/researchers participated in all the projects. Table 2 presents the partici- pants' profiles.

We cannot describe details about the projects due to confidentiality issues, but we can present some relevant characteristics. Digital Twin (DT) projects include the development of two tools to automate optimization processes and involve the analysis of data that is input to a model. In the Artificial Intelligence Projects (AI), we present four developed tools that involve the development of machine learning models, data analysis, and visualization. Logistic Projects (LOG) include two tools to automate logistics processes, data analysis, and visualizations; and the Process Mining Projects (PM) group developed one tool that promotes the automation of processes, data analysis, and visualizations. In general, all the teams performed data analysis, designed visualizations, and supported work in a complex domain, which involved understanding many specific concepts, data, and processes of the end users' routines. In Table 3, we characterize the projects evaluated in this research.

Table 2: Subjects' Profiles

| ID | Role | Exp in Sw Industry | Exp in UX | Exp in Data Science / ML | Project |
|---|---|---|---|---|---|
| P01 | Developer | 7 years | low | medium | DT |
| P02 | Developer | 3 years | none | medium | DT |
| P03 | Developer | 3 years | medium | medium | DT |
| P04 | Developer | 7 years | medium | medium | AI |
| P05 | Developer | 6 years | none | medium | AI |
| P06 | Developer | 9 years | medium | high | AI |
| P07 | Scrum Master | 10 years | low | medium | AI |
| P08 | Developer | 4 years | low | medium | AI |
| P09 | Developer | 7 years | medium | low | LOG |
| P10 | Developer | 12 years | low | low | LOG |
| P11 | Developer | 10 years | medium | low | LOG |
| P12 | Scrum Master | 28 years | medium | medium | LOG |
| P13 | Developer | 7 years | medium | high | LOG |
| P14 | Developer | 2 years | low | low | PM |
| P15 | Developer | 3 years | medium | medium | PM |
| P16 | UX/R | 7 years | high | medium | All |
| P17 | UX/R | 12 years | none | medium | All |
| P18 | UX/R | 20 years | high | medium | All |

For ethical reasons and to minimize the bias of participants becoming apprehensive because of the evaluation, all participants responded to an on- line consent form before answering the questionnaire. It made clear that the study did not evaluate the participants, their anonymity would be guaran- teed, and they could refuse to participate in the research without prejudice to their work. The researcher who analyzed the data was a member of the agile team. The other authors of this paper coordinate all projects. They did not know the participants' identification.

Table 4 presents the stages of the R&D Lean methodology that each project performed. The columns in the table represent the stages, as follows:

- 1 - Contextualization

Table 3: Project characteristics

| | DT | AI | LOG | PM |
|---|---|---|---|---|
| Machine Learning | | x | | x |
| Data Analysis | x | x | x | x |
| Visualization design | x | x | x | x |
| Automation of processes | x | | x | x |
| Complex domain | x | x | x | x |

- 2 - Lean Inception
- 3 - PBB
- 4 - Detailing UX and technical requirements
- 5 - Software Implementation
- 6 - Validation/User Testing
- user - stages in which users participated in the projects

Table 4: Stages and user participation in each project

| project | 1 | 2 | 3 | 4 | 5 | 6 | user |
|---|---|---|---|---|---|---|---|
| DT - tool 1 | X | X | X | X | X | X | all stages |
| DT - tool 2 | X | X | X | X | X | X | all stages |
| AI - tool 1 | | X | | X | X | X | stage 6 |
| AI - tool 2 | | X | X | X | X | X | all stages |
| AI - tool 3 | | | X | X | X | X | stages 4, 5, 6 |
| AI - tool 4 | X | | X | X | X | X | stage 6 |
| LOG - tool 1 | | | | X | X | X | stage 6 |
| LOG - tool 2 | | X | X | X | X | X | stages 2, 3, 4, 6 |
| PM - tool 1 | | | | X | X | | none |

### *3.2. Data Collection and Analysis*

We applied a questionnaire to collect data about the perception of the agile team regarding the participation of end users in the projects and the practices used both to understand and define the problem and to represent the solution. We used a questionnaire in this research due to the lack of availability of the agile team to participate in other activities that would de- mand more time. The questionnaire was composed of four modules aligned with the research questions: problem understanding (related to RQ1), solu- tion representation (related to RQ2), evaluating activities to understand the problem and to represent the solution (related to RQ3), and user participa- tion (related to RQ4). The questions were elaborated based on the results of previous works (Teixeira. et al., 2021; Ferreira et al., 2021). We sought to assess how much the points raised earlier were manifested in other projects. The questionnaire comprised open and closed questions (see Appendix A). Two other researchers reviewed the questionnaire, and the questions were grouped according to the research questions. We distributed the question- naire to the teams and gathered responses from 18 participants.

We analyzed both qualitative and quantitative results. One researcher used the codification process to analyze the answers collected in the open questions, and another researcher analyzed the created codes. The codes were associated with some themes defined previously: difficulties in understanding and defining the problem, practices to understand and define the problems, practices in representing the solution, difficulties in representing the solution, and participation of end users. First, these themes were identified for each project and for the UX and Research team. Finally, the codes were analyzed according to their similarities, in order to identify the final themes.

## 4. Qualitative Results

In the following subsections, we report the results associated to the teams of each project (Case 1, Case 2, Case 3, and Case 4) and the aforementioned themes. In Case 2, we uncovered a relevant theme that was not discussed in the others projects: difficulties to represent the solutions. Also, at the end of the qualitative results report, we present the perceptions of the UX and Research team that participated in all projects and thus had a broader overview.

### *4.1. Case 1: Digital Twin Project*

In this project, the team developed tools to optimize the processes of simulation software (Digital Twin) in the customer organization and a collaborative application. Three developers of this project answered the questionnaire.

Difficulties in understanding and define the problem: One diffi- culty in understanding and defining the problem identified in this project was the lack of specialists of the problem domain in the early ideation activities. This occurred in the second LI performed in this project: *"[In the Lean Inception] we lacked people who understood the problem in-depth to help determine the need and challenges."* - P01. The domain was specialized and involved several concepts that the team did not understand. Also, customers who were not end users did not know specific information: *"Due to the particular domain, sometimes not even [customers] understood the needs and the available data, which made it even more difficult for the (...) team to understand the problem."* - P01.

Practices to understand and define the problems: Respondents identified two practices that the team used to understand the domain and define the problem: customer meetings and the use of Lean Inception.

Customer meetings: *"(...) specific meetings where they [specialists] explained the functioning of the area of activity in which we are inserted"* - P02.

Use of Lean Inceptions: *"Lean Inception to understand the user needs"* - P01 and *"all documentation resulting from the Lean Inception"* - P03.

Regarding the Lean Inception organization, one respondent cited that developers did not participate in some activities, and they were idle during the process: *"there are activities where developers do not have to do much; it is more on the customer side"* - P03

Practices in representing the solution: Regarding the representation of the solution, we identified practices that the team adopted: prototype creation, solution architecture design, and meetings.

The architecture design represents the elements that compose the solution. It aids the development team in understanding the scope: *"A visualization of the architecture that we should develop was crucial, it changed over time, but it was always essential to have a clear view of where we needed to go"* - P02. The UX team designed the prototypes in two levels: low fidelity and high fidelity. The low-fidelity prototypes aided in the communication with the customers and end users. High-fidelity prototypes represented the final solution and aesthetics issues. In the question about which practices helped to represent the solution, the respondents cited: *"Prototyping/wireframe creation and meetings"* - P01.

End-user participation In this project, the end users participated in all stages of the project. Nevertheless, in some cases, they participated as observers and solved specific doubts, like in the second LI. From the results, we identified benefits of the participation of end users throughout the project and consequences of the customers acting as end users. The user participation benefits were: clarify needs and challenges, evaluate the developed product, solve doubts, and generate data for developers because the tools were based on data used in the processes. Some of the respondents' quotes were the following:

*"It contributed. The [end users] have helped a lot to clarify the needs and challenges."* - P01

*"Of course, because they are the ones who validate what we built"* - P03

*"It helped, as we had direct access to ask questions when necessary"* - P02

*"(...) they [end users] generated data that we needed to manipulate"* - P02

Regarding the consequences of the customers acting as end users, we identified customers that did not know the details of the users' routines: *" (...) sometimes not even [customers] understood the needs and the available data"* - P01. In this context, end users helped to solve some important doubts.

### *4.2. Case 2: Projects of Artificial Intelligence*

In this project, the team developed tools that use artificial intelligence to support activities and analysis in the routine of the operators (end users). The activities involved developing visualizations in dashboards and machine learning models. Four developers and the scrum master participated in this evaluation.

Difficulties to understand and define the problem: In the context of this project, we identified some difficulties in understanding and defining the problem: it is difficult to explain to customers the characteristics of solutions involving machine ;earning, the levels of uncertainty in the domain, and the number of features required to explain the phenomenon. Some quotes are as follows:

*"The difficulty in explaining the characteristics of Machine Learning solutions to the customer also creates confusion"* - P05

*"In Machine Learning projects, the uncertainty level of the application domain makes the understanding of the problem often unclear. Furthermore, the number of features that explain a phenomenon can also affect the understanding of the problem"* - P05.

Another difficulty encountered was that, in some cases, customers mediated the communication between the team and end users, and they did not know details of the users' routine: *"The 'customer' is unaware of the depth of business rules that the system must serve. Because we usually have as a customer someone from the managerial hierarchy, not from the area that will operate the system [end users], and whose goal is the business challenges, which are their day-to-day activities."* - P10.

Practices to understand and define the problem: Regarding the practices, we identified the following items:

Performing Lean Inception: *"Lean Inception conducted in [project name] helped a lot in understanding the problem"* - P05.

Analysis of existing solution: *"Solutions developed in the past and which today do not serve the business help understand the problem"* - P05.

Contact with end users: *"[about contact with users] It helped a lot. In [project 1], understanding the problem was more efficient than in [project 2]"* - P06 and *"End-user feedback helped us to understand the issues and improve the quality of the developed solution"* - P08.

Creation of prototypes: *"Paper prototyping, mock-ups, and interface versioning were important for aligning expectations and understanding requirements and changes during the interface design and development stages"* - P06

Sprint review meetings: *"results obtained in each sprint are frequently validated, which helps to identify whether the team is on the right path or not"* - P05.

Difficulties to represent the solutions: A difficulty cited in representing solution in this project was the lack of formal representation for the machine learning systems: *"Basically because it does not have a for- mal representation. I believe this is a problem with ML-based systems, which in turn are data-based"* - P05. Another issue was the end users' difficulty in explaining their needs regarding the interface: *"Often the user does not have any idea of how the system should behave. This is evident when they do low-level prototyping (e.g., using paper)"* - P06. A solution to minimize this problem was to create usage scenarios with users: *"At these times it is necessary to build together with them and elaborate usage scenarios so that* *they demonstrate how they expect the system to behave"* - P06.

Practices representing the solution The main practice cited in this project to support the representation of solutions was creating prototypes. Prototypes were important to support the discussions and to align the understanding of all stakeholders' profiles, and represent the final solution: *"The high-fidelity prototypes (especially in the generation of Power BI dashboards) were fundamental for developing these solutions. It supported the discussion of the final product and the equalization of the solution's understanding by several stakeholders."* - P10

To complement the use of prototypes, the team used other practices: card sorting to organize information in the interface and analysis of existing solutions that users were already using: *"we used card sorting to understand the users' mental model used as input on the organization of the options on the screen, to present similar systems to help design system process flows, icon design, preparation of menus, and options. It is important to understand what kind of software users used to influence the design of new systems"* - P06

End-User participation: In this project, various tools were developed, and end-user participation occurred in different ways. For some tools, users did not participate in the early stages (ideation, conception, requirements); they only validated the final version of the tool. In other cases, the end users participated in all stages of the development (from early stages to validation). A difficulty cited for involving the end users in those activities was the limited time availability. User participation helped the team to understand the data used in the implementation: *"The PO, who was also a user, was fundamental in giving insights about the scope of the data used for its processing. It reduced the development time"* - P06. Also, users aid in understanding features that add value and improvements to the software: *"Usually, end users have a clearer understanding of the problem and about what adds value"* - P05 and *"End-user feedback helped us to understand the issues and improve the quality of the developed solution."* - P08.

### *4.3. Case 3: Projects of Logistics Issues*

In this project, the team developed tools to automate and simulate processes related to logistical issues of the organization: The LOG - tool 1 and LOG - tool 2 (see Table 4). The LOG-tool 1 did not use the Lean Inception workshop because it was one of the first tools in the projects and need short time to start. Four developers and the scrum master participated in this research.

Difficulties in understanding and defining the problem: In this project, the team thinks that the lack of end users made it difficult to un- derstand the problem: *"Sometimes the requirements were not communicated clearly because the end user did not participate in the meetings. Due to the lack of direct contact with the end user, there were a lot of requests for changes throughout the development of the feature"* - P09. Another difficulty identified was the lack of data, and the customer (who was not the end user) did not know enough details to define the requirements: *"Lack of problem definition, such as not having the data to make the prediction, or the client not knowing how to do a calculation that he asks us to do in the system"* - P13. and *"The main reason for the difficulties was due to the lack of partic- ipation from the end users. Customers did not know all the business rules of the product"* - P11.

Practices to understand and define the problem: The practices used in this project to understand the problem were: Meetings with cus- tomers, meetings with the UX team, conducting a Lean Inception workshop (only in LOG-tool 2), and creating prototypes.

Meetings with customers: *"Review and planning meetings with the cus- tomer helped in passing on the knowledge of requirements"* - P09.

Meetings with the UX team: *"The graphical interface development in Adobe XD greatly facilitated the visualization and understanding of which was expected to the design and occasional meetings with the UX/UI team for explanations"* - P09.

Conducting a Lean Inception workshop: *"Lean Inception - Days of im- mersion with clients to understand the system's goals, as well as vocabulary and concepts of the client's business. Recognition of who is who and what each one can contribute to the client team"* - P10 and *"The Lean Inception process was fundamental for surveying the problems to be dealt with and achieving goals"* - P11. In tools for which there were no Lean Inceptions (LOG-tool 1), the problem was not appropriately defined: *"Some system modules did not have an LI workshop, and thus we did not have a complete understanding of the demand"* - P12. In the LI workshop (in LOG-tool 2) the team defined an overview of the problem and prioritization of features. Nevertheless, the subjects missed stages to discuss UX and technical issues: *"I missed steps focused on UX and IT infrastructure. And probably, these steps are defined near the end, or even in post-Lean Inception events"* - P10. In this project there was the same perception as in the Digital Twin project, some profiles of the team did not participate in some activities due their availability, and they were idle during the process: *"Lots of people/profiles participate every day. We could have segregated the days with subjects/profiles relevant to the day's activities"* - P10.

Creation of prototypes: *"The initial wireframes of the system screens were of vital importance for defining the requirements and developing the software"* - P12.

Difficulties in representing the solution: One difficulty that arose from the representation of solutions was implementing what the prototypes represent: *"resources and interactions that were possible in the prototype proved to be unfeasible or very complex to implement"* - P10. To solve this problem, the developers presented the graphical components toolkit for the UX team to use as a basis for the elaboration of prototypes: *"to present the graphic components toolkit used by the project to serve as an inspiration to the design"* - P10.

Practices for representing the solution: Practices cited to support the representation of solutions were to create prototypes, product backlog building an UX activities: *"prototypes in Adobe XD"* - P10 and *"development of prototypes"* - P11 and *"PBB and UX design with the customers"* - P13.

End-user participation: In this project, user participation occurred in different ways. In one of the developed tools (LOG-tool 1), users only participated in the final validation phase of the software, and in the LOG-tool 2, users participated in Lean Inception and UX activities. In LOG-tool 1, in which the end users participated only in the final validation of the tool and the team did not use LI, the following problems occurred:

Concepts were defined only after development: *"Definition issues were only noticed a little before the first deployment in production, during training with operators."* - P10.

Features were not clear: *"Sometimes features were not communicated clearly because the end user did not participate in the meetings. During the development, there were many change requests due to a lack of direct contact with the end user"* - P09. Nevertheless, in the LOG-tool 2 designed with user participation, their needs were clearer: *"It contributed because they spend the day-to-day reality in their work and the needs they have"* - P09.

### *4.4. Case 4: Project of Process Mining*

In this project, the team developed a tool to analyze the processes of the customer's organization, using process mining techniques. This project did not use Lean Inceptions because it was one of the first tools of the projects (similar to LOG-tool1). Two developers of this team participated in this research.

Difficulties to understand and define the problem: In this project, the problems identified in understanding and defining the problem were: lack of documentation and understanding the goal of the tool: *"I missed some kind of document describing the problem, expected results, description of the approach in general, etc. In other words, knowledge about the problem was generally passed on in an unstructured way over several meetings with colleagues and clients"* - P15 and *"Difficulties about how end users would use the tool if that analysis would bring value to them. Difficulties in how to better position elements on the interface to make users' work easier"* - P15.

Practices to understand and define the problem: The practices used in this project were: the creation of prototypes and meetings with customers and team: *"When I entered the project, I had a high-fidelity prototype, which helped me to understand the problem"* - P15 and *"Meetings with customers were essential for understanding the problem in general. For specific front-end development problems, meetings with the team and definition of requirements were essential to clarify doubts and understand what the client needed"* - P14.

Practices to represent the solutions: Also, to support the represen- tation of solutions, the developers cited the creation of prototypes and meet- ings: *"Team meetings, where we managed to organize activities for members to present the proposed solution productively"* - P14 and *"[to represent solutions] high-fidelity prototypes"* - P15.

End-user participation: In this project, there was no participation of end users. In this context, some difficulties encountered were: Rework features - *"Some features had to be dropped or revised. A cost that could have been avoided if you had talked to end users before implementing"* - P15 and customer mediating communication between user and team: *"In general, I think there was a lack of conversation and a better understanding about the end user, as the conversations were very focused on the customer. I think that difficulties occurred because the base for decisions was perceptions of the customer thought would be best for the users. The problem is that, in this case, the customer is not the user"* - P15

### *4.5. Perception of the UX and Research team*

The team consisted of two UX designers and a data science researcher. This team participated in all projects. We presented perceptions of this team about understanding/definition of the problem, the representation of solutions, and the participation of users in the projects.

Difficulties in understanding and defining the problem: In the projects of Digital Twin and Artificial Intelligence, the team had difficulties in understanding the domain due to their complexity and specific concepts: *"the (...) team was involved in specific meetings without having the knowledge and baggage necessary to monitor/make contributions"* - P16
*"(...) the domain is confusing and sometimes difficult to follow. [the customer organization] is more familiar and sometimes disregards that [technical team] is in a position of less knowledge on the subject."* - P16

*"Lack of understanding of the worked domain. I believe this happened, as I had never worked with the (...) area before"* - P17.

Another difficulty identified was that it was not always possible to validate the tools with end users: *"[Difficulty in understanding] Better context for using the tools and not always the possibility of validating the tool's efficiency with the end user"* - P18. Also, the UX team missed design-oriented activities in the projects: *"Need for ideation/design and visual references earlier in the process"* - P18.

Practices to Understand and define the problem: The UX and Research team considered that the practices that supported the understand- ing of the problem were: meetings with customers, performing Lean Incep- tions, PBB, UX/UI ideation, user interviews, and tests with users: *"Frequent meetings with our partners [customers] to understand, evolve, and eventually fix the proposed solutions"* - P17 and *"LI [Lean Inception], PBB, UX-UI ideation, user interviews, UTs [User Testing]"* - P18.

Practices to representing the solution: Regarding the practices to represent the solutions, the UX and Research team cited: meetings with experts in the domain documents explaining the variables(in projects that involve data science) and wireframes/prototypes: *"Meetings with experts and documents detailing the worked variables [in projects that involve data sci- ence]"* - P17 and *"wireframes and prototypes in high fidelity"* - P16.

End-user participation: Regarding user participation, the UX and Research team cited that, when the end users participated, they helped the team to understand their work routine and needs: *"[user participation] contributed to a better understanding of how they work, but at the same time, the domain is confusing and sometimes difficult to follow"* - P16 and *"The response from end users was significant to adapt the application to their actual needs."* - P17.

## 5. Quantitative Results

Ferreira et al. (2021) conducted interviews with customers and end users in a Digital Twin project, which also uses Lean Inceptions. They identified some issues regarding activities and artifacts used in the project. In our research, we created closed questions to evaluate the agreement of the agile team with the results identified previously.

Ferreira et al. (2021) reported that a workshop before the Lean Inception is necessary to help understand the problem domain and define the focus of the LI. Also, they identified that the agile team had guided the workshop before the LI to indicate the information that was important to explain in this activity. To evaluate the agreement of the teams with these issues, we added the following items in the questionnaire:

Q1. There must be a leveling of knowledge of the domain to be treated before the realization of Lean Inception so as not to hinder the progress of the workshop.

Q2. The development team (developers, designers, scrum master, POs) must recommend the information they deem necessary for the domain knowledge sharing to be performed before Lean Inception.

Only 14 out of the 18 respondents answered the questions about the Lean Inceptions activities because four of them did not participate in Lean Inceptions. Regarding Q1, ten respondents agreed that it is necessary to include a domain knowledge sharing session before the LI, three had a neutral opinion, and one disagreed. Regarding Q2, eight respondents agreed that the team should recommend the information that is essential to clarify at a session prior to the LI, five had a neutral opinion, and one disagreed. In this context, we observe that most of the subjects agreed with Q1. Regarding Q2, the quantity of agreements was close to the quantity of neutral and disagreement opinions. Figure 3 presents a summary of the results.

Regarding the users' information that is important for creating personas, most respondents considered that the role of the user, the profile and users' needs are important. Eight respondents considered that behaviors are important for personas creation. Other information cited as important in personas creation was user values, relationships, routine activities, and user problems. Figure 4 presents a summary of the results.

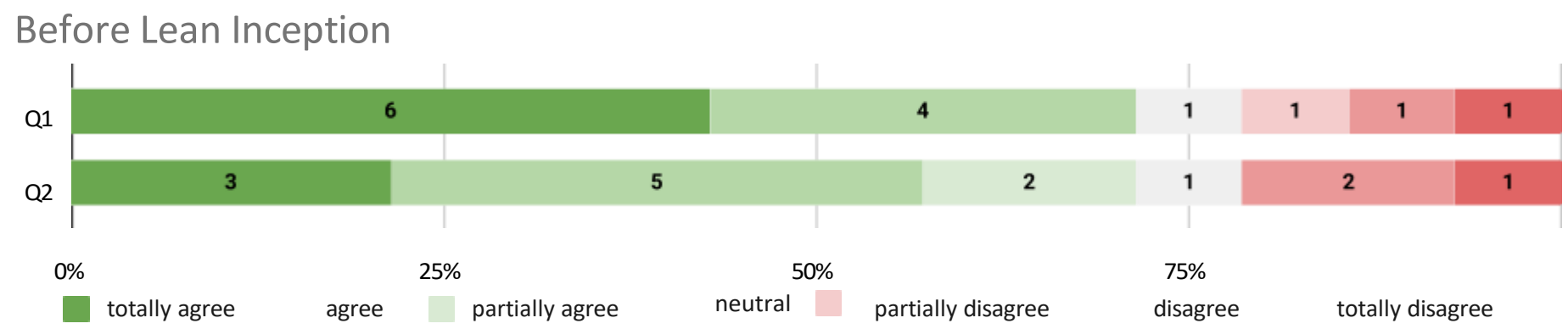

Figure 3: Activities before Lean Inceptions

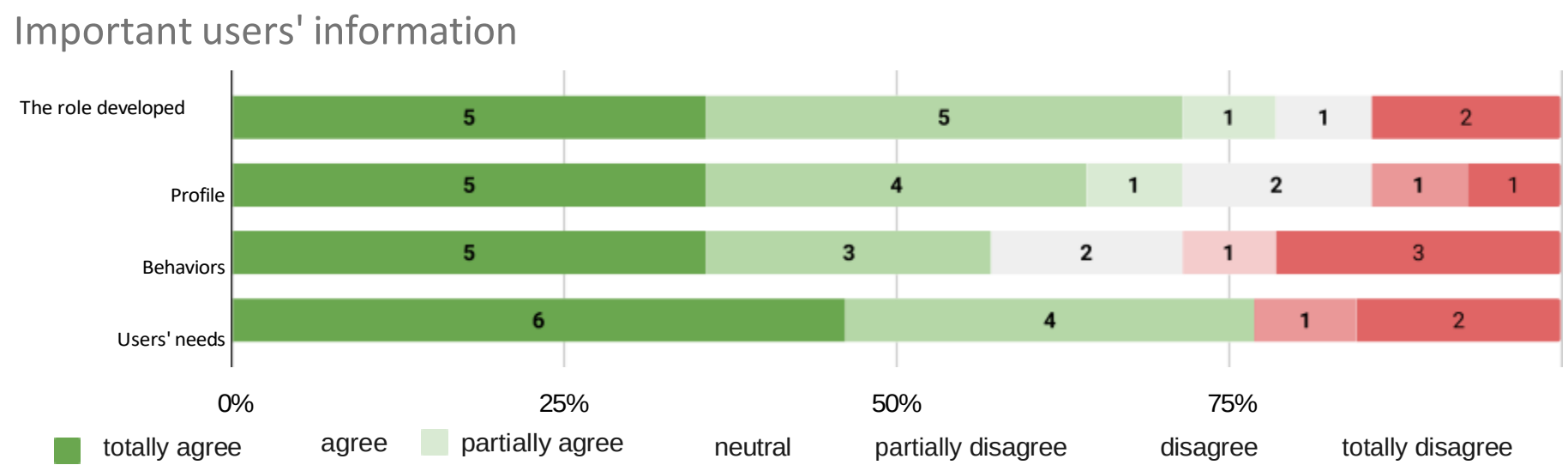

Figure 4: Asking users for relevant domain information

Additionally, we evaluated the respondents' agreement with the utility of user journeys for identifying features. Most of the subjects (ten out of 14) agreed with this issue. Regarding the Lean Inception, we asked whether it helps define what to do and not how to do it. As a result, the quantity of agreement (8 respondents) was close to the quantity of neutral and disagree- ment opinions (6 respondents). Another question was about the utility of the LI's activities to make the visualization of the problem less abstract. As a result, we identified that the activities that helped visualize the problem less abstractly were the product vision and brainstorming of features. According to most of the respondents, the activity that did not achieve this goal was the MVP canvas. Regarding the other activities (scoping, personas, user journeys, analysis of features values, and sequencing features), the quantity of agreement was close to the quantity of neutral and disagreement opinions. Figure 5 presents a summary of the results. Some respondents preferred not to opine and were not counted in the results.

Another issue evaluated was the usefulness of the creation of workflows. Figure 6 presents a summary of the results. Most of the subjects agreed that creating workflows helps identify elements to understand the problem, ma- terialize ideas discussed in Lean Inceptions, support wireframes/prototypes

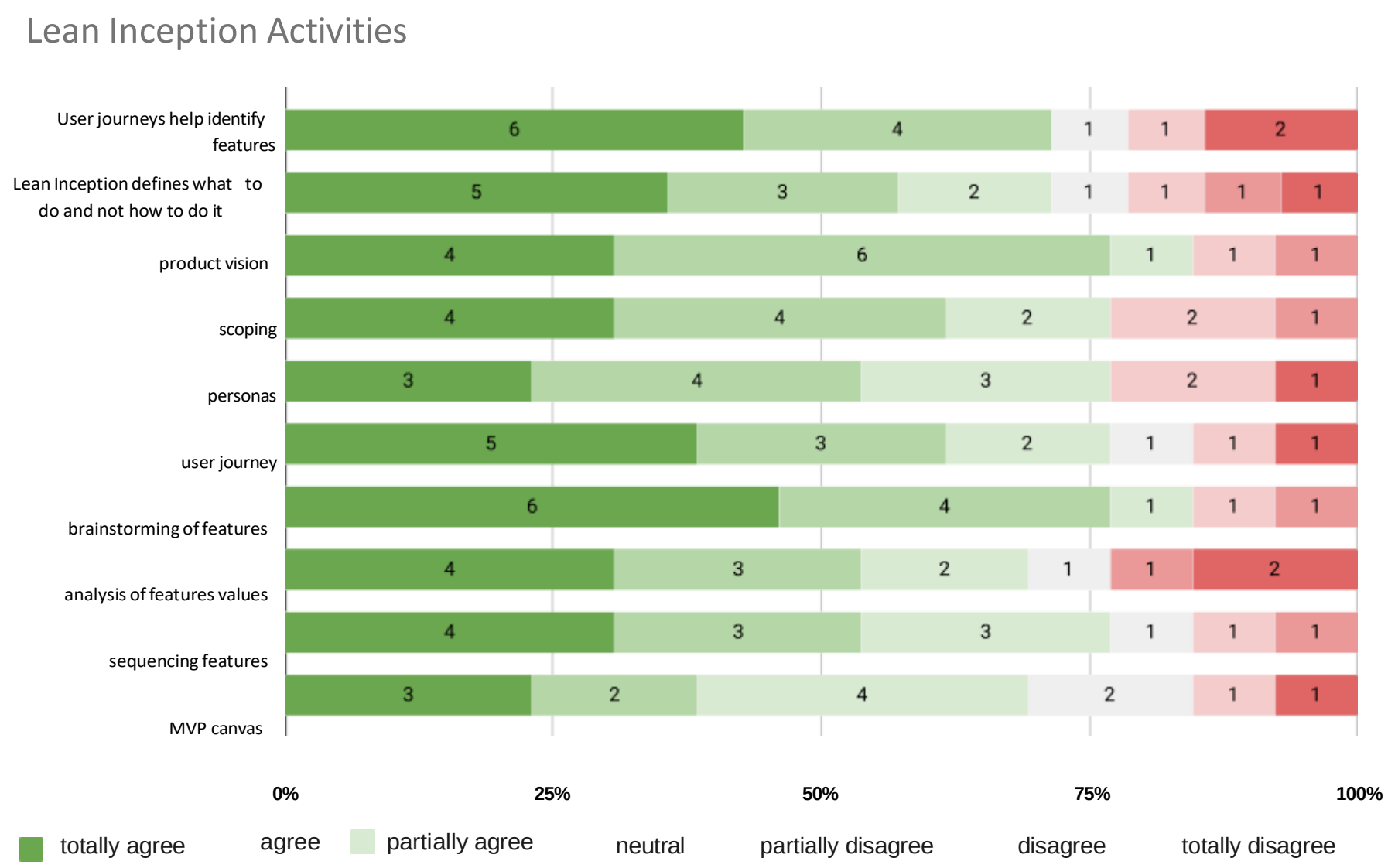

Figure 5: Lean Inception Activities

design, and converge on the idea of the solution.

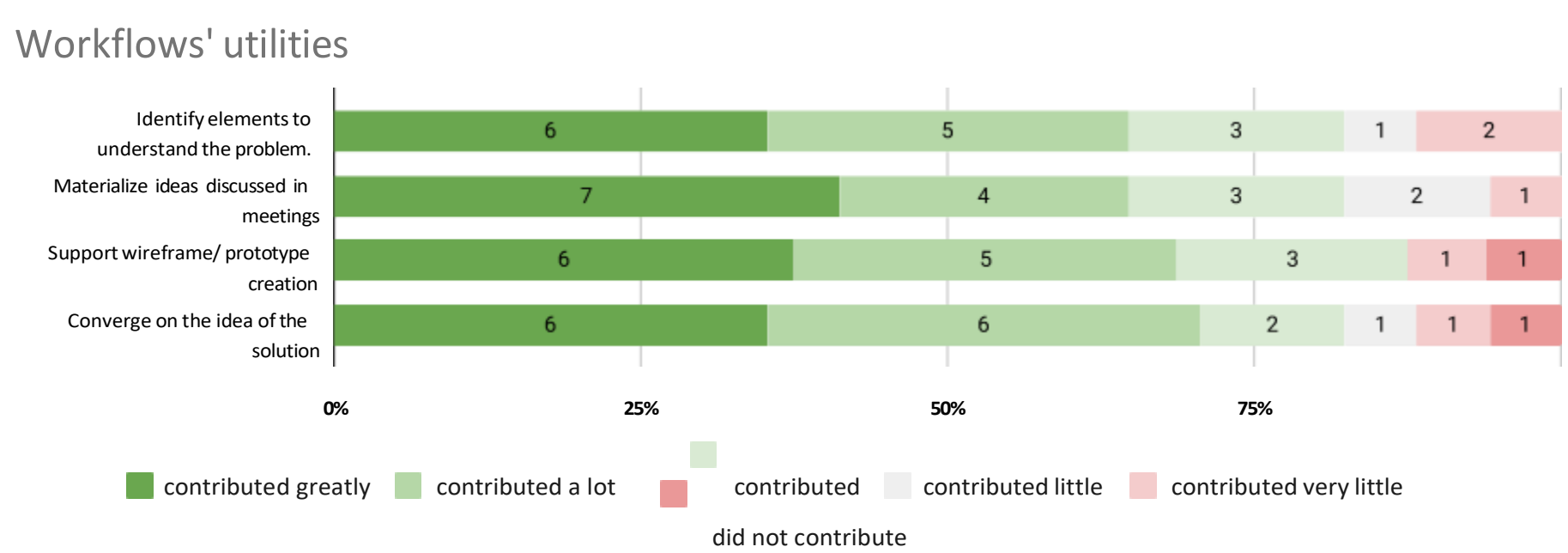

Figure 6: Workflows' utilities

Figure 7 presents the agreement of the respondents with the usefulness of the prototypes. Most of the respondents agreed with all the items evaluated. The items were: materialize ideas discussed in meetings, evolve/develop the ideas discussed, view the discussed concepts, facilitate discussions between those involved in the project Delimit the MVP, pass on knowledge to the development team, help capture ideas, and help understand calculations/rules to be implemented.

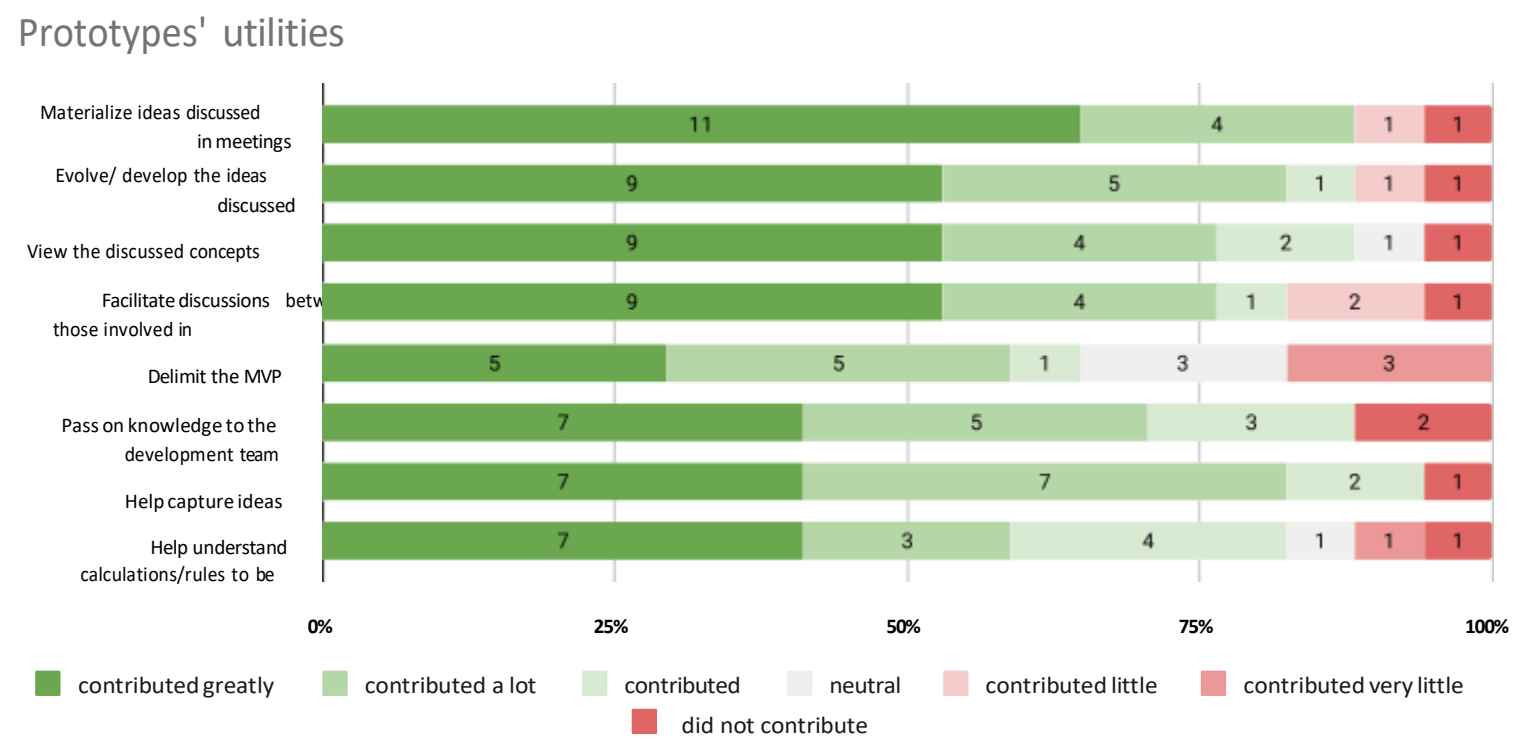

Figure 7: Prototype’s utilities

In the next item (see Figure 8), participants informed in which stages of the development they found that the users must participate to achieve some goals. The evaluated goals were:

- S1. Understand where and how to fit the product in the users’ routine
- S2. Reduce users’ resistance to include the product in their routine
- S3. Minimize usability problems
- S4. Increase perceived value in the use of the product
- S5. Identify barriers to implementation
- S6. Develop an easytouse product
- S7. Clearly define the problem
- S8. Align the problem with users’ needs

When users participate in the LI, the validations, and the tests, they integrate the software into their routine (S1). Users who mainly participate only in the final validations resist using the software in their routine (S2). Users should participate throughout development, validation, and testing to minimize usability issues (S3). Participants found that users who participate in the LI and validations and tests better perceive the product value (S4). For users to help identify barriers in implementation (S5), they must participate in all stages equally, except the product backlog activity and the early stages.Most respondents agreed that users must be present throughout the software development to help identify barriers to implementation (S6). The stages of the development where participants felt that user participation would be most important to clearly define the problem (S7) and align the problem with users' needs (S8) were the early stages of the process and the LI.

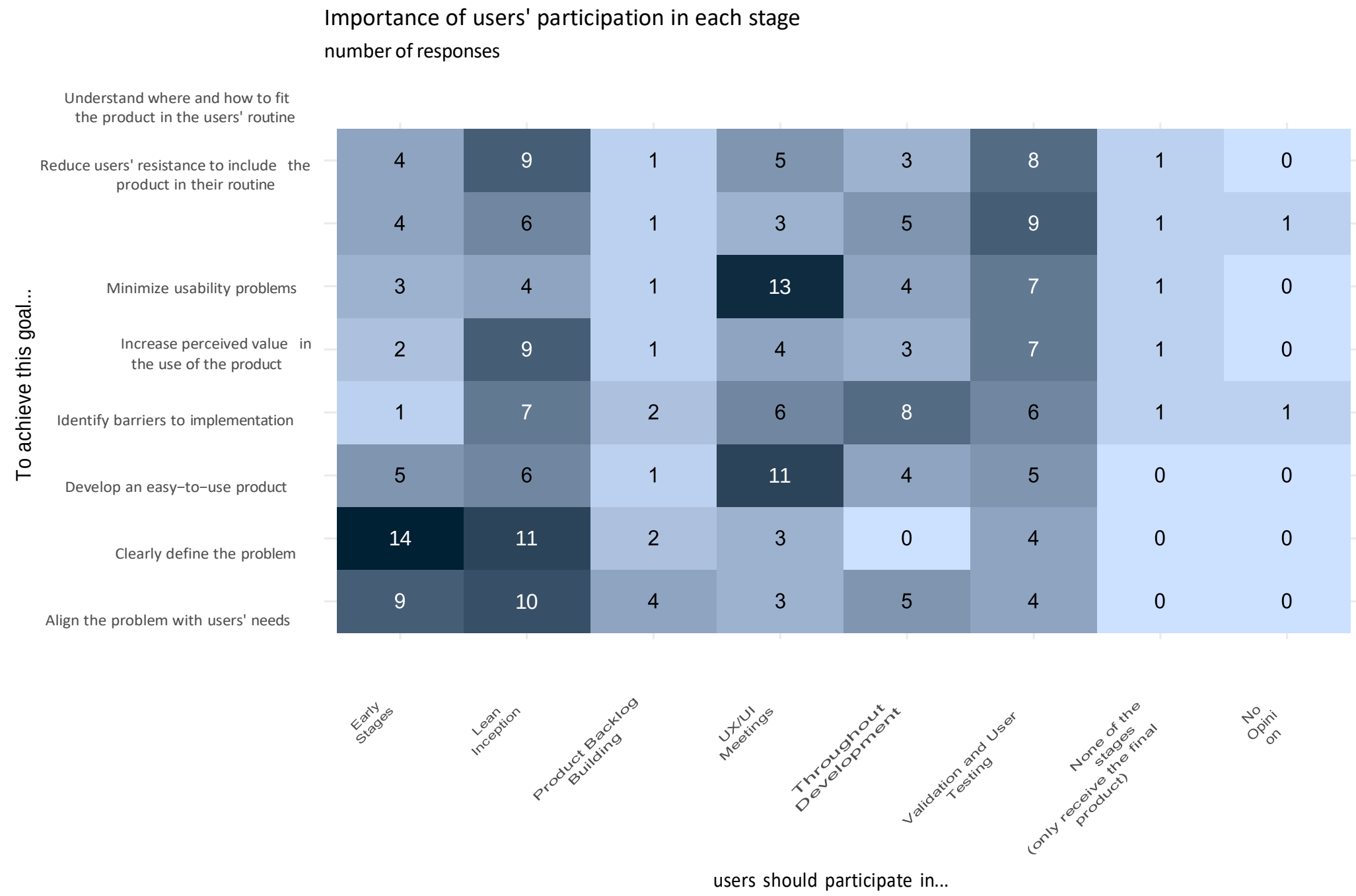

Figure 8: Users' participation issues

## 6. Discussion and Lessons Learned

In this paper, we present a qualitative (summarized in Figure 10) and a quantitative study to investigate which practices were used by four agile teams to support the understanding and definition of the problem (RQ1) and the representation of solutions (RQ2), which activities presented diffi- culties (RQ3), and the perceptions about end-user participation (RQ4) in the projects. We now summarize the main issues that influenced the design and development of the projects and answer the research questions.

### *6.1. Q1. Which practices supported understanding and defining the problem in these types of projects?*

Regarding the understanding of the problem, we identified that the practices that supported this step in most projects were the realization of Lean Inceptions and the elaboration of prototypes. The LI's were important for understanding the general goals of each project, and the prototypes supported the joint understanding of the users' needs. The LI's were important to understand the general goals of each project. However, in data science projects, there are many uncertainties because they involve data analysis and understanding. Also, in projects that involve process automation, it is necessary understanding specific concepts. In this context, post-LI activities were necessary to complement the understanding of the problem and data, and resolved processes.

The project's second most often used practice was holding meetings with stakeholders (end users and customers). From the quantitative analysis, another practices that was considered useful to view the concepts an under- standing the problem were the user journey, personas and workflows. We identified some information relevant to create personas and help in under- standing the problem (see Figure 6: the role developed, profile, behaviors and users' needs.

In the quantitative analysis, most of the participants agreed that the team needs to have a minimal level of knowledge about the domain before the LI, and the team must recommend the important information to improve this initial knowledge. These issues are presented in Figure 5.

### *6.2. RQ2. Which practices supported representing the solutions in these types of projects?*

The artifacts used to represent the solutions in the projects were mainly prototypes and meetings with customers, where doubts were clarified. Also, from the quantitative analysis, the creation of prototypes was considered a useful practice in various aspects: they materialize ideas discussed in meetings, evolve/develop the ideas discussed, allow the team to view the discussed concepts, facilitate discussions between those involved in the project, delimit the MVP, pass on knowledge to the development team, help capture ideas, and help understand calculations/rules to be implemented.

In cases involving AI projects, due the quantity of features to explain the phenomena, even with end-user participation, it was necessary to use other methods besides prototypes to facilitate the communication with users to understand the problem. In this case, the team created usage scenarios before prototyping.

### *6.3. RQ3. Which difficulties were encountered in defining and understanding the problem and in representing solutions in these types of projects?*

In LI, the teams missed stages to discuss UX and technical issues. Also, the developers did not participate in some activities. They were only ob- servers when the activity did not motivate their participation or when, in some cases, they could not follow the explanations about the unfamiliar do- main.

Another issue that made it difficult to understand and define the problem was the complexity and uncertainty of domains that involve data analysis and understanding of various processes in the organization. For example, there were many features to explain certain phenomena in the AI projects, which made it difficult to understand the problem. In the context of these projects, post-LI activities were necessary to complement the understanding of the problem and resolved processes.

In projects that involved implementation using Machine Learning, we identified a lack of formal representation for the developer to use as a basis. One of the developers, who joined one of the projects later, mentioned that, due to the lack of documentation, it was difficult to understand the problem addressed by the project. He missed some form of structured documentation.

Another difficulty was the technical viability. The UX team created high-fidelity prototypes based on the low-fidelity prototypes already developed in a co-creation process with the stakeholders. Nevertheless, in some cases, the implementation of the prototypes was not technically viable or was too costly. A solution for this problem was to present the development toolkit elements for the UX designers to use as inspiration to create prototypes.

One of the problems identified in these practices was that the end user did not always participate directly in activities of design and decision making. In many cases, customers mediated the communication between the agile team and end users. Nevertheless, they would not use the final product and were not aware of the challenges and needs that arose in the end users' work routines. Other problems that arose due to the lack of user participation in the initial stages of ideation and development were that problems with incorrect understanding of domain concepts were identified after the development stages, during training with users.

### *6.4. RQ4. How was the end-user participation in these projects?*

Users were very important in projects that involved analyzing data and implementing machine learning models. They helped the development and UX team understand the available data and the types of data analysis that would bring value to their work. In projects that automated processes, users helped understand those processes and what was technically feasible to implement. In all projects where users participated directly, they were critical in helping to define business rules that were not known in depth by customers. Figure 9 presents the perceived benefits of end-user participation in each project group.

### *6.5. Lessons Learned*

In these projects, we identified the following lessons learned:

1 - Lean Inceptions were important to understand the general goals of data science and process automation projects. However, due to the uncertainties and complexity of their domains, post-LI activities were necessary to complement the understanding of the concepts, data, and processes. LI fits in the area of 'concrete experiences with users' present work' because it aids the team in understanding the users' work. This area was proposed in the model of communication presented in (Kensing and Munk-Madsen, 1993). LI provides Software Engineering with a sample of what exists in PD and co-design in other works. We present evidence that it is viable to explore these activities better in the software engineering field in this context.

2 - End-user participation is important because they understand the problems in their work better than anyone else. They help the development and UX team understand the available data, the types of data analysis that would bring value to their work, understand work processes, and what was technically feasible to implement.

3 - When customers mediate the communication between the agile team and end users, domain concepts can be incorrectly understood. It occurred because they would not use the final product and were unaware of the challenges and needs that arose in the end users' work routines.

4 - In projects that involved machine learning, we identified a lack of formal requirements representation for the developer to use as a basis.

5 - Developers present the development toolkit elements for the UX designers to use as inspiration. It aims to avoid the implementation of prototypes that are not technically viable or too costly for development.

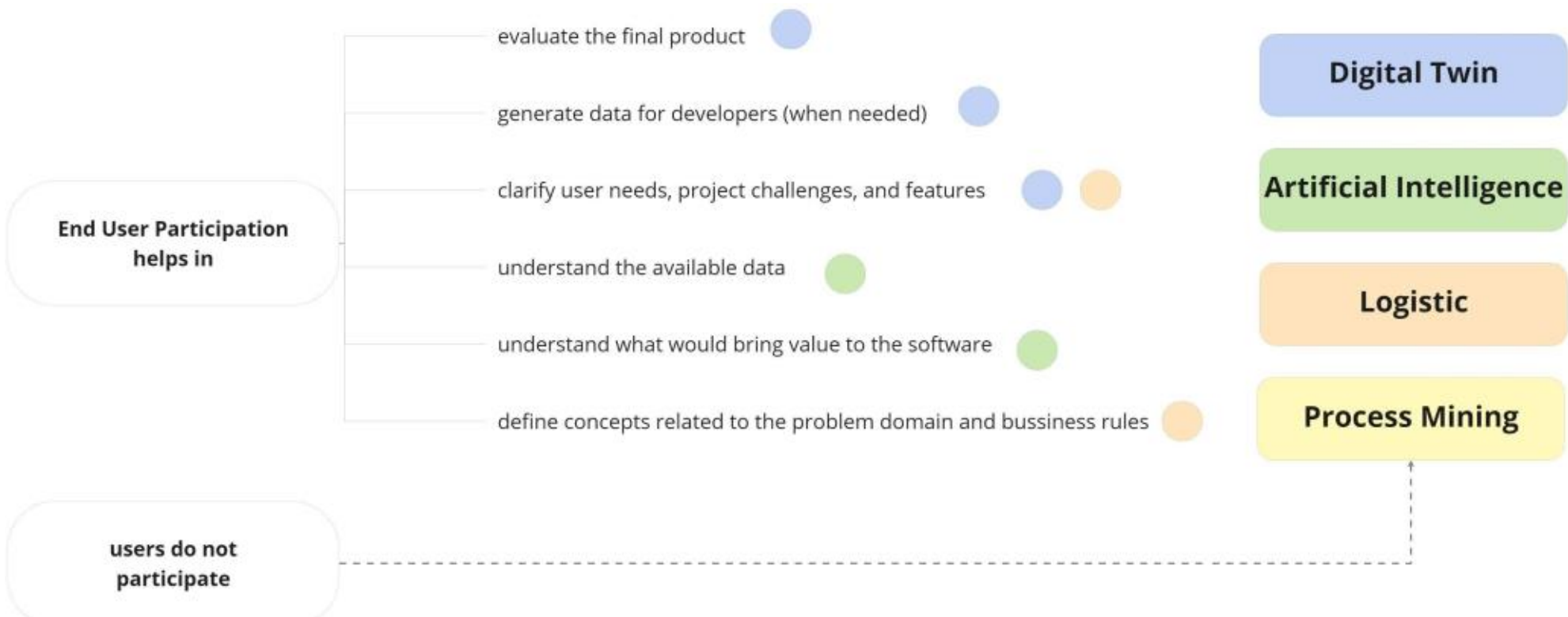

Figure 9: Benefits of end-user participation

## 7. Threats to Validity

Every study possesses threats that can affect the validity of their results (Wohlin et al., 2012). For questionnaire-based studies, face validity, content validity, criterion validity, and construct validity should be discussed together with reliability concerns Kitchenham and Pfleeger (2008) *apud* Linåker et al. (2015).

Face validity. This validity refers to subjectively evaluating the survey understanding. It is typically mitigated by conducting lightweight reviews of the questionnaire by randomly chosen respondents. After the creation of the questionnaire by one researcher, two other researchers revised it to avoid ambiguities, inconsistencies, and lack of clarity in the questions.

Content validity. This validity refers to the subjective evaluation of how appropriate the instrument seems to reviewers experts in the subject matter. This type of validity is typically mitigated by conducting reviewers with subject matter experts. The two researchers that reviewed the questionnaire were experts in the topics of UX and software engineering. Besides face validity, they also focused on confirming that the questions were in line with the research goals.

Criterion validity. This type of validity refers to how the questionnaire can separate between respondents that belong to different groups. It is mitigated by determining which groups an instrument should identify. We included questions to enable precisely separating the respondents by the projects in which they participated.

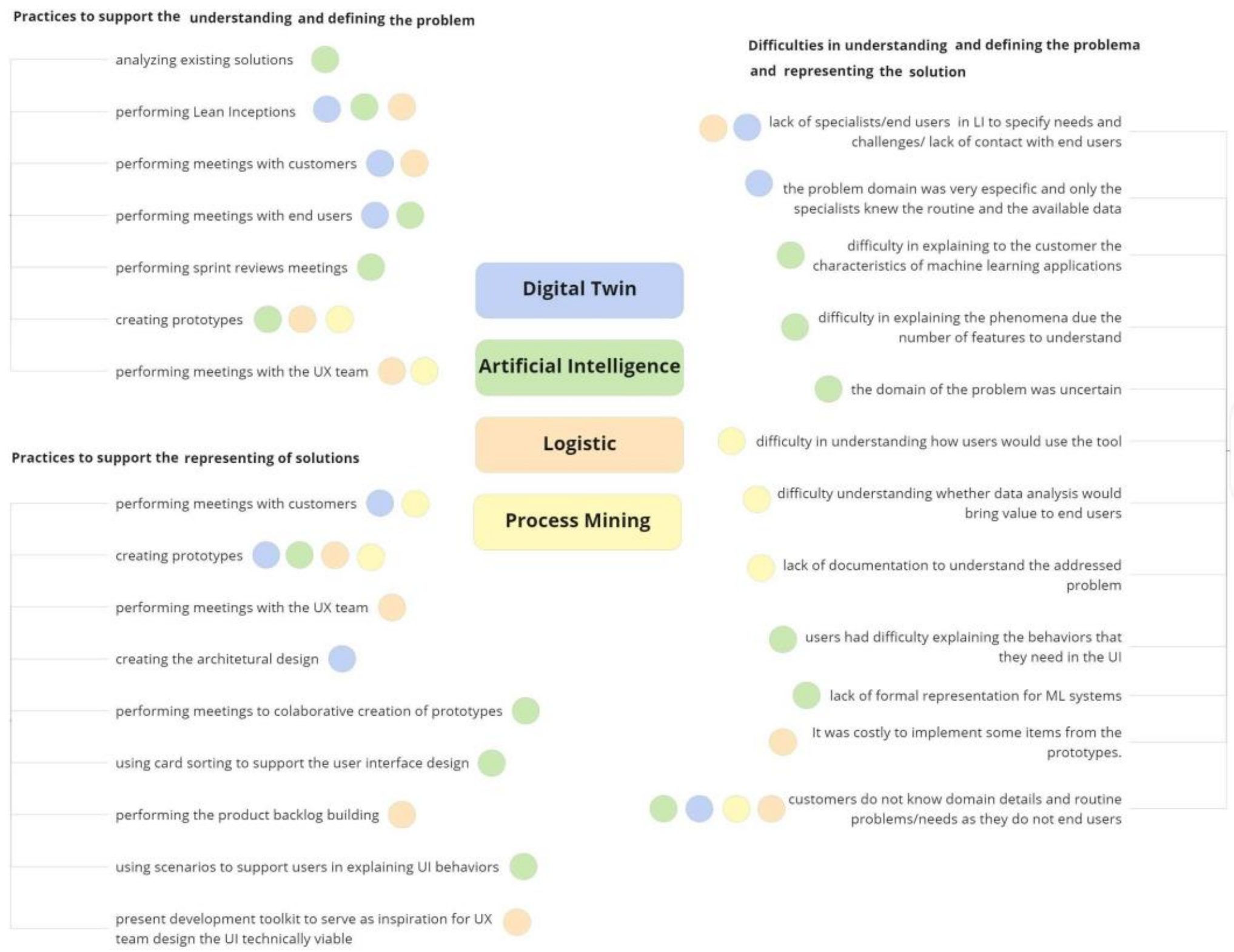

Figure 10: Practices and difficulties in understanding and defining the problem (top) and in supporting the representation of solutions (bottom)

Construct Validity. This validity refers to how well an instrument mea- sures the construct it is designed to measure. The questionnaire used for data collection was designed to address the research questions. The specific questions and answer options were based on observations gathered in previ- ous interview-based investigations in other projects (Teixeira. et al., 2021; Ferreira et al., 2021).

Reliability. It refers to reliability and consistency of the conclusions de- signed. To avoid researcher bias and improve the reliability of the results, as suggested by Wagner et al. (2020), we used independent validations. The qualitative analyses were conducted by one researcher and the results ana- lyzed by two others. For confidential reasons, raw data was not made avail- able. Regarding generalizability, while we had subjects from four different agile teams, the data gathered in this research represents the opinions and experiences of the participants and may not necessarily be generalizable to other development contexts. Maxwell (2012) defined the internal generalizability as ”the generalizability of a conclusion within the case, setting, or group studied, to persons, events, times, and settings that were not directly observed, interviewed, or otherwise represented in the data collected”. Our research covers all profiles developers, scrum master/PO, UX designers and researchers (see Table 2 of these projects to address this threat. Also, We highlight that the most of the team participated in this research: 13 out of 22 developers, 2 out of 3 scrum masters/PO and all UX designers and researchers. Some team members had already left from the projects when this study was carried out.

Nevertheless, we consider that this research revealed practices and lessons learned that could be relevant to other projects, in particular those involving data science and process automation.

## 8. Conclusions and Future Work

In this paper, we presented a study in the context of an academia-industry collaboration. The study was performed with four agile teams in projects that involve data science and process automation. We investigate three issues in this context: the problem space, the solution space, and the end-user participation.

Regarding the problem space, that is, the practices and difficulties to understand and define the problem, we identified that the practices most used were prototypes and meetings with stakeholders. A problem, in this case, was that too often the stakeholders were customers and not end users; therefore, they did not know the problems in the users’ routine. In this context, it was challenging to define needs, goals, and business rules. Specifically, in the context of data science projects, the participation of end users helps in understanding the data and which analyses would deliver most value to them. Another practice that helped the team understand the problem was the ideation workshop: the Lean Inception (LI).

Loi et al. (2019) cite some challenges and opportunities to integrate AI and HCI. One of the challenges is “it is difficult for designers to imagine synergy with technical ML”, and some opportunities were “UX designers working together with technical data scientists can lead to promising outcomes” and “HCI+AI collaborations can help make systems more usable, but also spark new AI innovations." In this context, our research not only brings further evidence of the importance of such collaboration, but also discusses ways in which this collaboration may take place, contributing to the practice of engaging UX designers, data scientists, and ML developers in collaborative projects.

Regarding the solution space, we identified that the developers missed a formal representation of requirements to support the implementation of the features. Also, the UX team needed to find the graphical elements of the development toolkit that would allow them to create technically viable prototypes. The end users play an important role in the implementation stage because they provide the team with the data and identify barriers in the work routine that affect the solution development (for instance, specific data formats or the lack of certain data).

Finally, we observed that the end users had an important role in the development process and that the agile team shared this perception. Nevertheless, we lacked strategies and concrete approaches to improve end-user participation in the ideation and development activities in agile projects. As future work, we will apply the lessons learned in this research to improve our process and conduct additional studies to evaluate those improvements.

## Appendix A. Questionnaire

1. Problem understanding questions

    1.1. What requirements/information/artifacts/practices were essential for a better understanding the project's problem? In what cases were they necessary?

    1.2. During the project, did have difficulties in understanding the needs of the stakeholders (customers, end users, or others) or the worked problem?

    1.3. What difficulties arose in understanding the problem? How/Why did these difficulties arise?

2. Solution representation questions

    2.1. What requirements/information/artifacts/practices were important to support representation (design/development) of the solution? In what cases were they important?

2.2. During the project, were there any difficulties in representing the solution?

2.3. What difficulties arose in the representation (design/development) of the solution? How/Why did these difficulties arise?

2.4. What actions were taken by the team to address the difficulties in understanding mentioned in the previous question?

3. Evaluating Activities to understand the problem and to represent the solution

Lean Inception

3.1. In which projects do you participate in Lean Inception?

3.2. About Lean Inception, describe what you thought was good and what could be improved?

3.3. Check how much you agree with each of the statements below: [Options: 1 - strong agree; 2 - agree; 3 - weak agree; 4 - neutral; 5 - weak disagree; 6 - disagree; 7 - strong disagree]

- There must be a leveling of knowledge of the domain to be treated before the realization of Lean Inception to prevent the progress of the workshop.

- Development team (developers, designers, scrum master, POs) must recommend what information they deem necessary for the leveling (mentioned in the previous item) to be performed before Lean Inception.

3.4. Check how much you agree with each of the statements below: [Options: 1 - strong agree; 2 - agree; 3 - weak agree; 4 - neutral; 5 - weak disagree; 6 - disagree; 7 - strong disagree]

- User journeys help identify features
- Lean Inception defines what to do and not how to do it

3.5. Check how much you agree that each of Lean Inception's activities helped to materialize/visualize (make less abstract) the understanding of the problem to be addressed: [Options: 1 - strong agree; 2 - agree; 3 - weak agree; 4 - neutral; 5 - weak disagree; 6 - disagree; 7 - strong disagree] - product vision
- scoping
- personas
- user journey
- brainstorming of features
- analysis of features values
- sequencing features
- MVP canvas

3.5.1. Other comments about Lean Inception activities

## Personas

3.6. Check how much you agree that the following user characteristics were important to detail during Lean Inception: [ Options: 1 - strong agree; 2 - agree; 3 - weak agree; 4 - neutral; 5 - weak disagree; 6 - disagree; 7 - strong disagree]

-The role developed
-Personal Profile
-Behaviors
-Users' needs

3.6.1. What other characteristics about users do you think are essential to address in the activities?

## Workflows

3.7. Check how much the elaboration of flows (mapping the steps to achieve the users' tasks) contributes to the following items:

[Options: 1- contributed strongly; 2 - contributed a lot; 3 - contributed; 4 - neutral; 5 - contributed little; 6 - contributed weakly; 7 - did not contribute]

- Identify elements to understand the problem
- Materialize ideas discussed in meetings
- Support wireframes/ prototype creation
- Converge on the idea of the solution

3.7.1. Other comments about using flows

Prototypes

3.8. Check how much the elaboration of user interface prototypes contributes to the following items: [Options: 1- contributed strongly; 2 - contributed a lot; 3 - contributed; 4 - neutral;5 - contributed little; 6 - contributed weakly; 7 - did not contribute] - Materialize ideas discussed in meetings
- Evolve/ develop the ideas discussed
- View the discussed concepts
- Facilitate discussions between those involved in the project
- Delimit the MVP
- Pass on knowledge to the development team
- Help capture ideas
- Help understand calculations/rules to be implemented

3.8.1. Other comments about the elaboration of user interface prototypes

4. User Participation

4.1. In the project(s) you worked/worked on, was there contact with end users?

*If 4.1 is an affirmative answer:*

4.2. In which phase(s) of the projects in which you participated did you contact users? [project-name - phase]

4.3. Did the contact with end users contribute/hinder the understanding of the problem or the design of the solution? How and why?

*If 4.1 is a negative answer:*

4.4. Did the lack of contact with end users contribute/ hinder any activity? How and why? If so, how were they resolved/overcome?

*If there were difficulties[in 4.4]:*

4.5. How were they resolved/overcome?

4.6. Check which stage(s) listed above about the participation of end users are important to achieve the following goals (you can have more than one option if you want):

Stages:
1 - From the beginning of the process;
2 - At Lean Inception;
3 - In the construction of the backlog (PBB);
4 - At design/UX/UI meetings;
5 - Throughout development;
6 - In the evaluations/tests of the final product;
7 - Do not participate, only receive the final product

Goals:
- Clearly define the problem
- Check the stages: [1][2][3][4][5][6][7]

- Align the problem with users' needs
- Check the stages: [1][2][3][4][5][6][7]

- Increase perceived value in the use of the product
- Check the stages: [1][2][3][4][5][6][7]

- Develop an easy-to-use product
- Check the stages: [1][2][3][4][5][6][7]

- Reduce users' resistance to include the product in their routine
- Check the stages: [1][2][3][4][5][6][7]

- Understand where and how to fit the product in the users' routine
- Check the stages: [1][2][3][4][5][6][7]

- Minimize usability problems
- Check the stages: [1][2][3][4][5][6][7]

- Identify barriers to implementation
- Check the stages: [1][2][3][4][5][6][7]